\documentclass[iop,revtex4]{emulateapj}
\usepackage{amsmath}
\usepackage{amssymb}
\usepackage{graphics}
\usepackage{graphicx,epsfig}
\usepackage{pdflscape}
\usepackage{tabularx}
\usepackage{natbib}
\usepackage{datetime}
\usepackage{morefloats}
\usepackage[export]{adjustbox}
\shorttitle{Fan Loops}
\shortauthors{Ghosh et al.}

\begin{document}
\title{Fan Loops Observed by IRIS, EIS and AIA}
\author{Avyarthana Ghosh\altaffilmark{1,2}, Durgesh Tripathi\altaffilmark{1}, G. R. Gupta\altaffilmark{1}, Vanessa Polito\altaffilmark{3},
Helen E. Mason\altaffilmark{3}, Sami K. Solanki\altaffilmark{4,5}}
\affil{$^1$Inter-University Centre for Astronomy and Astrophysics, Post Bag - 4, Ganeshkhind, Pune 411007, India}
\affil{$^2$Center of Excellence in Space Sciences India, Indian Institute of Science Education and Research, Kolkata, West Bengal 741246, India}
\affil{$^3$Department of Applied Mathematics and Theoretical Physics, Wilberforce Road, Cambridge CB3 0WA, UK}
\affil{$^4$Max-Planck Institute for Solar System Research, Justus-von-Liebig-Weg 3, 37077 G\"ottingen, Germany}
\affil{$^5$School of Space Research, Kyung Hee University, Yongin, Gyeonggi-Do, 446-701, Korea}

\email{avyarthana@iucaa.in}
\begin{abstract}
A comprehensive study of the physical parameters of active region fan loops is presented using the observations recorded with the Interface Region Imaging Spectrometer (IRIS), the EUV Imaging Spectrometer (EIS) on-board Hinode and the Atmospheric Imaging Assembly (AIA) and the Helioseismic and Magnetic Imager (HMI) on-board the Solar Dynamics Observatory (SDO). The fan loops emerging from non-flaring AR~11899 (near the disk-center) on 19th            November, 2013 are clearly discernible in AIA 171~{\AA} images and those obtained in \ion{Fe}{8} and \ion{Si}{7} images using EIS. Our measurements of electron densities reveal that the footpoints of these loops are approximately at constant pressure with electron densities of $\log\,N_{e}=$ 10.1 cm$^{-3}$ at $\log\,[T/K]=5.15$ (\ion{O}{4}), and $\log\,N_{e}=$ 8.9 cm$^{-3}$ at $\log\,[T/K]=6.15$ (\ion{Si}{10}). The electron temperature diagnosed across the fan loops by means of EM-Loci suggest that at the footpoints, there are two temperature components at  $\
log\,[T/K]=4.95$ and 5.95, which are picked-up by IRIS lines and EIS lines respectively. At higher heights, the loops are nearly isothermal at $\log\,[T/K]=5.95$, that remained constant along the loop. The measurement of Doppler shift using IRIS lines suggests that the plasma at the footpoints of these loops is predominantly redshifted by 2{--}3~km~s$^{-1}$ in \ion{C}{2}, 10-15~km~s$^{-1}$ in \ion{Si}{4} and $~$15{--}20~km~s$^{-1}$ in \ion{O}{4}, reflecting the increase in the speed of downflows with increasing temperature from $\log\,[T/K]=4.40$ to 5.15. These observations can be explained by low frequency nanoflares or impulsive heating, and provide further important constraints on the modeling of the dynamics of fan loops.
\end{abstract}


\section{Introduction}
The observations from modern high resolution instruments reveal that active regions comprise a variety of loop structures. These loops are considered to be the building blocks of the solar corona. Therefore, a comprehensive understanding of the physics of all kinds of loops is key to the problem of solar coronal heating \citep[see][for reviews]{Kli_2006, Rea_2014, MooB_2015, Kli_2015}.

Active region loops are broadly classified into three categories {--} namely, hot core loops (3{--}5~MK), warm loops (1{--}2~MK) and fan loops (0.6{--}1 MK). In addition, there is a significant amount of diffuse plasma spread over a large area at coronal temperatures without any well-defined visible structures, possibly due to the absence of instruments with sufficiently high spatial resolution \citep{DelM_2003,ViaK_2011,SubTK_2014}. 

The hot loops are rooted in moss regions \citep{BerPF_1999, AntKD_2003, TriMD_2010, TriMK_2012} and have electron densities $\log\,N_{e}$ = 9.58~cm$^{-3}$ and 9.26~cm$^{-3}$ for \ion{Fe}{14} ($\log\, [T/K]=$ 6.30) and \ion{Fe}{13} ($\log\, [T/K]=$ 6.25) respectively \citep{BroDT_1997,TriMD_2010,Del_2013}. The observations of hot loops reveals that a range of frequencies of heating events may be present in the core of active regions \citep{WarBW_2011, TriKM_2011, WarWB_2012, DelTM_2015}. Warm loops are believed to be multi-stranded structures with electron densities ranging between $\log\,N_{e}=$ 8.5 to 9.0 cm$^{-3}$. Their properties can be explained by low frequency impulsive heating \citep[see e.g.,][]{DelM_2003, WarWM_2003, Kli_2006, TriMD_2009, UgaWB_2009, GupTM_2015}.


The fan loops are possibly the most complex and longest living loop structures, formed at the periphery of active regions and were first studied in detail by \cite{SchTB_1999} using the observations recorded by the Transition Region and Coronal Explorer \citep[TRACE;][]{HanAK_1999}. These are thought to be rooted in the penumbrae of sunspots, in close proximity to active regions with relatively strong magnetic fields, in enhanced network zones or even in unipolar quiet network regions. Such structures connect regions of large flux concentration over distances as large as $10^{5}$~km or more \citep{SchTB_1999}. While their lifetimes typically range from several hours to days, the evolution time scale is a fraction of an hour \citep{SchTB_1999}.  

One of the earliest spectroscopic studies of fan loops was performed by \cite{WinHB_2002} using observations made with the Solar Ultraviolet Measurements of Emitted Radiation \citep[SUMER;][]{WilCM_1995} on-board the Solar and Heliospheric Observatory (SoHO). These fan loops were observed in the emission line of \ion{Ne}{8}~770~{\AA} ($\log\, [T/K]=$ 5.80) and the plasma in these loops showed persistent downflows (redshifts) of 15-40~km~s$^{-1}$. We, however, note that the reference wavelength used to derive the Doppler shift was 770.409~{\AA} and that was revised to 770.428$\pm$0.007~{\AA} by \citet[][]{PetJ_1999} and to 770.428$\pm$0.003~{\AA} by \cite{DamWC_1999} using SUMER observations. The use of the revised wavelength will therefore reduce the Doppler shift in fan loops computed by \cite{WinHB_2002} by $\sim$5-10~km~s$^{-1}$. Based on hydrodynamic modelling, the observed flows were attributed to non-uniform asymmetric heating of the loops. Later on, \citet{MarWX_2004} studied fan loops observed over 
three active regions and found Doppler velocities of $\sim \pm 5$~km~s$^{-1}$ for \ion{H}{1} Ly$\beta$~1025~{\AA} ($\log\, [T/K]=$ 4.00)  and $\sim \pm 2$~km~s$^{-1}$ for \ion{Si}{2}~1533~{\AA} ($\log\, [T/K]=$ 4.20). At higher temperatures, the redshifts increased to $\sim 5$~km~s$^{-1}$ in \ion{C}{4}~1548~{\AA} ($\log\, [T/K]=$ 5.05), and $\sim 15-20$~km~s$^{-1}$ in the spectral lines of both \ion{N}{5}~1548~{\AA} and \ion{O}{6}~1031~{\AA} formed at $\log\, [T/K]=$ 5.30 and 5.45 respectively. However, the redshift decreased to $\sim 10$~km~s$^{-1}$ in the spectral line of \ion{Ne}{8}~770~{\AA}. However, \citet{Dos_2006} reported that the plasma flowing along the field lines in these fan loops was blueshifted by 5-10~km~s$^{-1}$ in \ion{Ne}{8}~770~{\AA} and \ion{S}{5}~786~{\AA} ($\log\, [T/K]=$ 5.20) lines.

\begin{figure*}
\centering
\includegraphics[width= 0.9\textwidth]{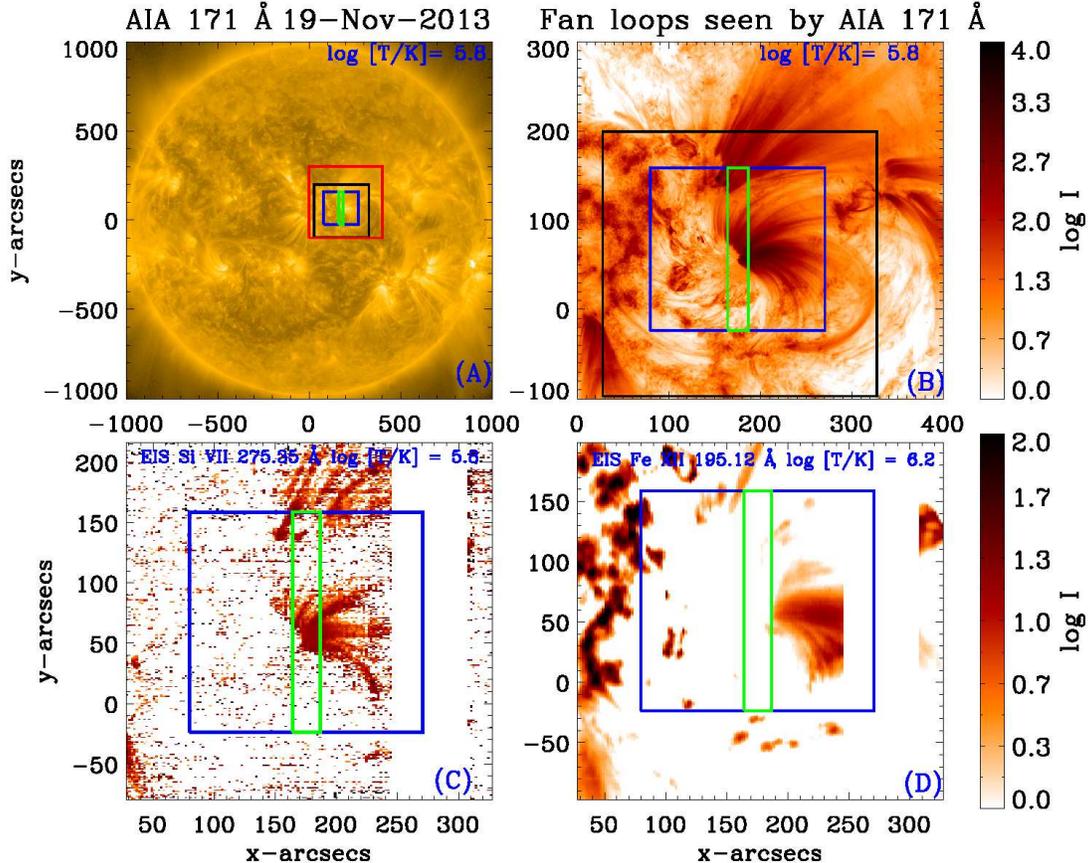}
\caption{\textit{(A)}: Full-disk AIA image at 171~{\AA} showing AR 11899, recorded on 19 November 2013 at 10:39:59 UT with the over-plotted red box being the AIA FOV zoomed-in in panel (B).  \textit{(B)}: AIA 171~{\AA} image (in negative brightness scale) showing the fan loops. \textit{(C)}: EIS \ion{Si}{7} 275.35~{\AA} raster image (in negative brightness scale) showing the fan loops between 10:40:20 UT and 11:59:00 UT. \textit{(D)}:  EIS raster image of the same region in the \ion{Fe}{12} 195.12~{\AA} line (in negative brightness scale). In (A) and (B), the black box is the EIS raster FOV (CCD B). In all four panels, the superimposed blue and green boxes represent the IRIS Slit-Jaw image (SJI) and IRIS raster FOV at 10:36:39 UT, respectively.}\label{context}
\end{figure*}

With the launch of the EUV Imaging Spectrometer \citep[EIS; ][]{CulHJ_2007} on-board Hinode, the measurements of physical parameters such as electron densities, temperatures and Doppler shifts in various structures within a temperature span from the upper transition region to the corona have been routinely performed \citep[see e.g.,][]{MarWW_2008, Del_2008, DosWM_2008, TriMD_2009, TriMK_2012, DadTT_2012, WinTM_2013}. Doppler shifts of the plasma confined in fan loops were measured using EIS observations by \citet{WarUY_2011} and \citet{YouDM_2012}. \citet{WarUY_2011} showed that the plasma at the footpoints of fan loops was redshifted by $\sim$~30~km~s$^{-1}$ in \ion{Si}{7} ($\log\, [T/K]=$ 5.80) line and also suggested, based on magnetic field extrapolation, that fan loops are closed loop structures, though the other footpoints may not be visible in coronal images. \citet{YouDM_2012} reported that plasma in the fan loops was redshifted ($\sim~15-20$~km~s$^{-1}$) in \ion{Fe}{8} line at $\log\,[T/K] =$ 5.80, 
but were blueshifted ($\sim 25$~km~s$^{-1}$) in the emission lines of \ion{Fe}{12} (above $\log\,[T/K]$ = 6.20) at their footpoints. At intermediate temperatures (\ion{Fe}{10} line, $\log\,[T/K]= $ 6.00), they observed mixed signatures of downflows and upflows. 
 
The Interface Region Imaging Spectrograph \citep[IRIS;][]{DePTL_2014}, which was launched in 2013, when combined with the EIS, provides a remarkable opportunity to study the various physical plasma parameters in the solar atmosphere all the way from the chromosphere to the corona. In this paper, we study a set of fan loops emanating from a sunspot using simultaneous observations recorded by IRIS, EIS, the Atmospheric Imaging Assembly \citep[AIA;][]{LemTA_2012} \& the Helioseismic and Magnetic Imager \citep[HMI;][]{SchBN_2012,SchSB_2012}, on-board the Solar Dynamics Observatory \citep[SDO;][]{PesTC_2012}. The rest of the paper is structured as follows. In \S\ref{obs}, we provide a brief description of the instruments used, and discuss the processing techniques. Analysis and results are presented in \S\ref{res}, followed by a summary and discussion of the results in \S\ref{con}.

\section{Observations} \label{obs}
\begin{table*}[t]
\caption{IRIS and EIS spectral lines used for studying the fan loops emanating out of AR 11899 on 19th November 2013, where $\lambda_{0}$ is the rest wavelength. The peak formation temperatures are taken from CHIANTI \citep{DerML_1996, LanYD_2013} at one particular density.}\label{spectral_lines}
\begin{center}
\begin{tabular}{| c | c | c | c | c | c |} \hline

\multicolumn{1}{|c}{} &\multicolumn{1}{c}{\textbf{IRIS Lines}} &  &\multicolumn{1}{c}{} &\multicolumn{1}{c}{\textbf{EIS Lines}} &  \\
 
\multicolumn{1}{|c}{} &\multicolumn{1}{c}{} & &\multicolumn{1}{c}{} &\multicolumn{1}{c}{} &  \\ \hline
 
Ion name   &$\lambda_{0}$    & Peak   &Ion name  &$\lambda_{0}$    & Peak \\ 
			& \citep{SanB_1986}  &formation  &  &\citep{BroFS_2008}            &formation \\
             &              &temperature   &          &           &temperature \\ 
             & ~{(\AA)}        & (log) (T/K)          &          &~{(\AA)}           & (log)(T/K)\\ \hline
\ion{C}{2} 			&1334.532 &4.40	&\ion{Fe}{8}		& 194.663 & 5.65\\
\ion{C}{2} 			&1335.708 &4.40 & \ion{Si}{7}		& 275.368  & 5.80\\
& & & &\citep{{WarUY_2011}} & \\
\ion{Si}{4}			&1393.755 &4.90 & \ion{Si}{10}$^{a}$& 258.375 & 6.15\\
\ion{Si}{4}			&1402.770 &4.90 & \ion{Si}{10}$^{b}$& 261.058 & 6.15\\      
\ion{O}{4}$^{a}$	&1399.755 &5.15 & \ion{Fe}{12}  	& 195.119 & 6.20\\
\ion{O}{4}$^{b}$ 	&1401.156 &5.15 & \ion{Fe}{13}  	& 202.044 & 6.25\\ 
            		&     	  &    	& \ion{Fe}{14}		& 264.787 & 6.30\\ \hline
\end{tabular}
\end{center} 
{~~~~~~~~~~~~~~~~~~~~~~~~~~~~~~~~~~~~~~~~~~~~~~~~~~~~~~~~~~~~~~~~~~~~~~~~~$^{a,b}$\textit{density sensitive line pair}} \\
\end{table*}

\begin{figure*}[t]
\centering
\includegraphics[width=0.9\textwidth]{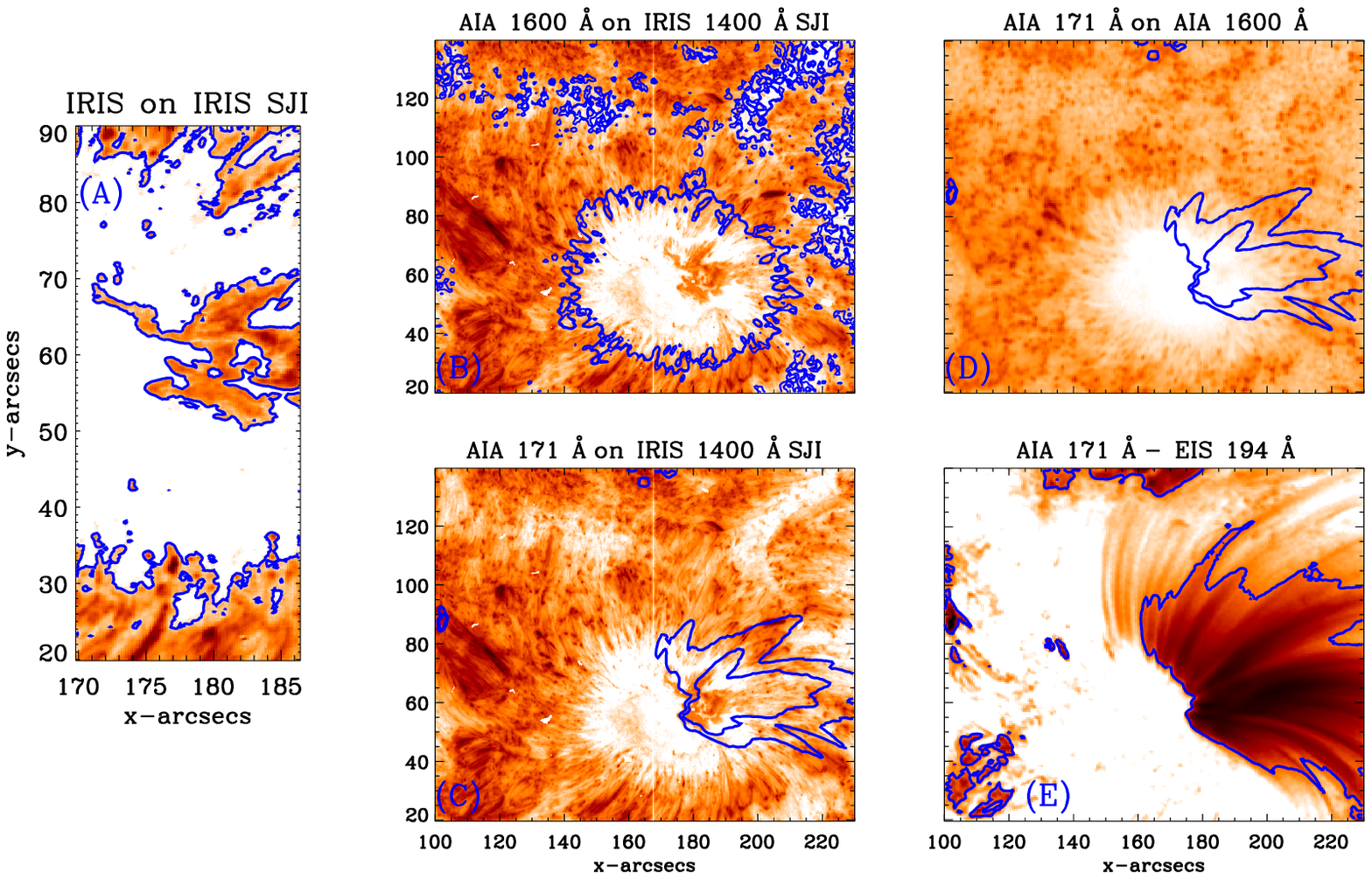}
\vspace{-5mm}
\caption{Co-aligned images of AIA, EIS and IRIS at comparable temperatures. \textit{(A)}: IRIS raster \ion{Si}{4} in the 1402.77~{\AA} spectral line (contours) co-aligned with IRIS \ion{Si}{4} 1400~{\AA} SJI (background). \textit{(B)}: AIA 1600~{\AA} image (contours) co-aligned with IRIS \ion{Si}{4} 1400~{\AA} SJI (background). \textit{(C)}: IRIS \ion{Si}{4} 1400~{\AA} (background) slit-jaw image co-aligned with AIA 171~{\AA} (contours). \textit{(D)}: AIA 171~{\AA} fan loops (contours) superimposed on AIA 1600~{\AA} (background) with co-alignment. \textit{(E)}: EIS \ion{Fe}{8} 194.66~{\AA} (contours) superimposed on AIA 171~{\AA} image (background) after co-alignment. The fan loop contours are clearly visible in the images (C), (D) and (E). All images display negative intensities.}\label{coalign}
\end{figure*}

The analysed active region (\textsl{AR~11899}) appeared on the east limb of the Sun on November 12, 2013 and was observed at heliographic coordinates of 284W, 31N on the 19th of November, 2013. On that date, this active region was observed nearly simultaneously by Hinode/EIS, IRIS, SDO/AIA~\&~HMI. Figure~\ref{context}~(A) displays the full-disk AIA image taken with the 171~{\AA} channel. The over-plotted black box is the EIS raster field-of-view (FOV) for CCD B, the blue box is the IRIS Slit-Jaw Image (SJI) FOV and the green box is the IRIS raster FOV. The red box in figure~\ref{context}(A) is the region highlighted in figure 1(B) showing the fan loops in detail. The EIS raster images of the active region using two emission lines (\ion{Si}{7} 275.35~{\AA} at $\log\,[T/K]=5.80$ and \ion{Fe}{12}~195.12~{\AA} at $\log\,[T/K]=6.20$) are plotted in figures~\ref{context}~(C) \&~(D). The maps in  figures~\ref{context}~(B), (C) \&~(D) are plotted using a negative intensity scale. The gaps in Figures~\ref{context}~(C)
 \&~(D) between x= [228\arcsec,292\arcsec] are caused by missing data in the EIS raster. Henceforth, this region has been neglected in our analysis. The blue and green boxes on these EIS intensity maps show the FOV of IRIS SJI and raster respectively. However, in all the later figures, the IRIS raster FOV have been reduced so as to focus on the footpoint region only.
 
To determine the plasma densities and temperatures in the fan loops, spectroscopic data from IRIS and EIS have been utilised. For Doppler velocities, however, only IRIS observations are used. For this particular observation, EIS used the 2$\arcsec$ slit to raster over 150 positions ($i.e.$ time-steps) between 10:40:20~UT and 11:59:00~UT with an exposure of $\sim$30 seconds so that the EIS FOV is [300\arcsec ,300\arcsec]. IRIS rastered a FOV of [20\arcsec ,182\arcsec] six times over a period of $\sim$ 32 minutes (between 10:31:15~UT and 11:03:31~UT). Each raster is 5 minutes and 17 seconds long. The spectral lines, from IRIS as well as EIS, used for this study are listed in Table~\ref{spectral_lines}, along with their laboratory wavelengths taken from \citet{SanB_1986} for IRIS lines \& from \citet{BroFS_2008} for EIS lines. Note that the reference wavelength for \ion{Si}{7} mentioned in \citet{BroFS_2008} should be corrected to 275.368~{\AA} \citep[see][]{WarUY_2011}. The peak formation temperatures have 
been taken from CHIANTI \citep{DerML_1996, LanYD_2013}. 

In this study we have used level-2 data from IRIS and level-0 data from EIS. The IRIS data are corrected for all instrumental effects such as flat-fielding, dark currents and offsets, so as to make them suitable for all scientific purposes\footnote{A User’s Guide To IRIS Data Retrieval, Reduction $\&$ Analysis, S.W. McIntosh, February 2014} including the thermal orbit variations. Also, during the length of each raster of 5 minutes 17 seconds long, the orbital error should be negligible. However, we estimated the residual orbital variations for a single raster as well as over the entire duration of the six rasters and concluded that it is negligible. The IRIS data are analysed using Gaussian fitting routines provided in solarsoft\footnote{Using EIS Gaussian fitting routines for IRIS data, P. Young, April 2014}. EIS level-0 data have been pre-processed  with the eis\_prep.pro\footnote{EIS Software Note No. 13, P. Young, 2010} routine. For the wavelength calibration, orbital drift and slit tilt errors are two 
major sources of concern. The eis\_auto\_fit.pro routine\footnote{EIS Software Note No. 16, P. Young, 2015} rectifies the EIS spectral data by removing these errors. 

In this study we have data in two IRIS spectral windows, namely \ion{C}{2} and \ion{Si}{4}. Within the \ion{C}{2} window there are two \ion{C}{2} lines at 1334.5~{\AA} \& 1335.71~{\AA} ($\log\,[T/K]=4.40$), but the signal strength is poor for both of them. Hence, a $4\times4$ pixel binning is performed. The \ion{Si}{4} window harbours two lines at 1394.78~{\AA} and 1402.77~{\AA} ($\log\,[T/K]=4.90$). The \ion{Si}{4} window centered at 1402.77~{\AA} also has two \ion{O}{4} lines observed at 1399.77~{\AA} and 1401.16~{\AA} ($\log\,[T/K]=5.15$). For the \ion{O}{4} lines, due to poor signal-to-noise ratio (SNR), a $4\times4$ pixel binning is required. These two lines of \ion{O}{4} are density sensitive and are used for the measurement of electron densities under the assumption of a Maxwellian distribution of electron velocities \citep[see however,][]{DudDK_2011}. 

For the EIS spectral analysis (see Table~\ref{spectral_lines}), we use lines from \ion{Fe}{8} ($\log\,[T/K]=5.65$) to \ion{Fe}{14} ($\log\,[T/K]=6.30$). The \ion{Fe}{8}~194.66~{\AA} line has a blend in its red wing, at 194.80~{\AA} \citep{YouZM_2007}, which is removed using a double Gaussian fit. The \ion{Fe}{12} line at 195.12~{\AA} has a self-blend at 195.18~{\AA}, but its contribution is negligible ($<$10\%) in regions with densities lower than $\log\,N_{e} = 9.5$ cm$^{-3}$ at $\log\,[T/K]= 6.20$. Therefore, fitting a single Gaussian suffices. 
  
One important aspect of measuring the Doppler shifts is to determine a reference wavelength. Generally, neutral or singly ionized photospheric or chromospheric lines serve the purpose of determining the in-flight absolute wavelength drift \citep{HasRO_1991} when there are no calibration lamps on-board\footnote{IRIS Technical Note 20: Wavelength Calibration, January 9, 2013}. IRIS has a \ion{S}{1} line with a rest wavelength of 1401.5136~{\AA} \citep{DePTL_2014}. The observed wavelength of the same \ion{S}{1} line is 1401.52~{\AA} (averaged over the entire raster) which translates to a velocity difference of $\sim$1.0~km~s$^{-1}$. This line is used for the absolute wavelength calibration of all IRIS lines. For the EIS instrument, however, there are no neutral spectral lines or on-board calibration lamp. A method to obtain absolute velocities from EIS was derived by \cite{YouDM_2012} that used quiet Sun region in \ion{Fe}{8} line to obtain the reference wavelength. Unfortunately, in our observations, no such 
region could be identified. Hence, we have not attempted to derive Doppler velocities using EIS lines.

Our aim is to study the various physical parameters of the plasma within the fan loops using IRIS and EIS. Therefore, we need to co-align the EIS and IRIS images. Since AIA gives full-disk images at different temperatures, these can be used as references to co-align the IRIS and EIS observations. For this purpose, firstly the IRIS raster obtained in \ion{Si}{4}~1402.77~{\AA} (plotted in contours in figure ~\ref{coalign}(A)) is over-plotted on an IRIS Slit Jaw Image (SJI) taken in \ion{Si}{4}~1400~{\AA} (the reference image in (Figure ~\ref{coalign}(A))) to check for any misalignment. The \ion{Si}{4} SJI is then co-aligned with 1600~{\AA} images taken by AIA (Figure ~\ref{coalign}(B)). This is followed by co-aligned images of AIA 171~{\AA} channel on IRIS \ion{Si}{4}~1400~{\AA} SJI (Figure ~\ref{coalign}(C))) in background. Furthermore, AIA 171~{\AA} channel image has been co-aligned with AIA 1600~{\AA} image. The raster image obtained in EIS \ion{Fe}{8} was co-aligned with AIA images taken at 171~{\AA}. 
Figure~\ref{coalign} displays the co-aligned IRIS, EIS and AIA images.

\section{Data Analysis and Results} \label{res}
\begin{figure*}
\centering
\includegraphics[width=0.95\textwidth]{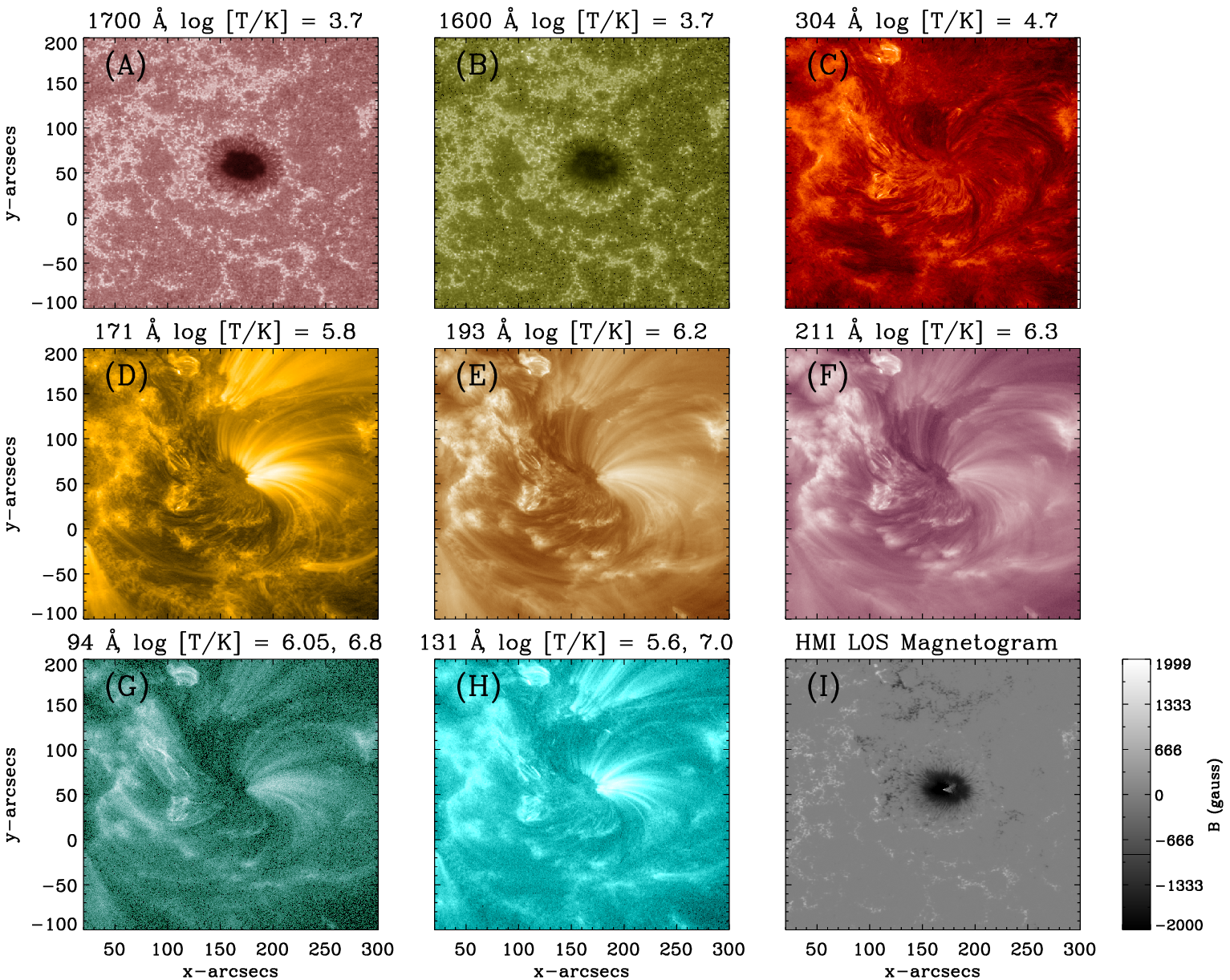}
\caption{Images of AR 11899 on 19-Nov-2013 in the 8 AIA/SDO channels in order of increasing temperature and a HMI LOS magnetogram. The channels and their corresponding peak formation temperatures (in log scale) are also noted.}\label{aia_context}
\end{figure*}

In Figure~\ref{aia_context}, the 1600~{\AA} and 1700~{\AA} images correspond to near-continuum, which show the sunspot umbra fringed by the penumbra and scattered bright plage. The 304~{\AA} channel primarily corresponds to the emission in the \ion{He}{2} line that shows the sunspot (which does not appear dark in this channel) and the active region in the hottest part of the chromosphere ($\log\, [T/K]= 4.70$). The images in the second and third rows of the Figure~\ref{aia_context} (D-H) display the morphology of the fan loops emanating from the sunspot at different characteristic temperatures. Figure~\ref{aia_context}(I) shows the line of sight (LOS) magnetogram. The magnetogram clearly indicates a bright region, corresponding to apparent opposite polarity field within the sunspot umbra, although this may well be a location of anomalous polarisation rather than true opposite polarity.


As can be seen in Figure~\ref{aia_context}, the fan loops are seen in almost all the channels of AIA. This is essentially due to the fact that all the channels have some contribution from low temperature lines forming below a million degrees \citep[see e.g.,][for more detail]{DwyDM_2010}. The fan loops are most prominent in the image taken in the AIA 171~{\AA} channel at $\log\,[T/K]=5.80$. As the temperature rises further, the loops become less and less perceptible. A similar effect was observationally seen for the warm loops by \citet{TriMD_2009} and modelled by \citet{GuaRP_2010}. However, the 131~{\AA} channel of AIA has a significant contribution from the \ion{Fe}{8} line formed at $\log\,[T/K]=5.60$. This \ion{Fe}{8} line emission actually exhibits itself as the very bright loops emanating from the footpoint region even in the 131~{\AA} channel. The 94~{\AA} channel has contributions from several transitions of \ion{Fe}{10} and \ion{Fe}{14} formed over a wide temperature range \citep{DelSB_2012}. The 
intermediate temperature channels, 193 and 211~{\AA} show gradual fading of the loops in the background whereas the cores are still visible. Examining the intensity maps obtained using EIS lines over a range of temperatures ($\log\,[T/K]=$ 5.65 to 6.30)  show that the loops are clearly discernible in the lower temperature lines and gradually fade at higher temperatures.

\begin{figure*}
\centering
\includegraphics[width=0.95\textwidth,left]{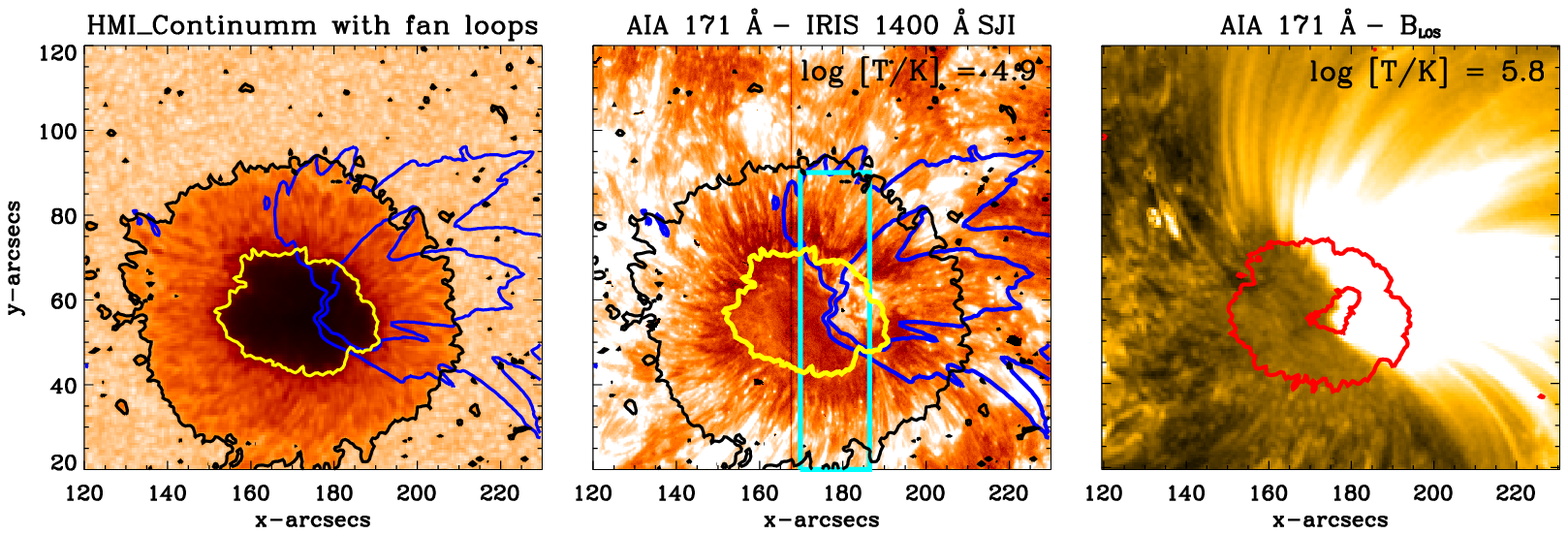}
\vspace{-35mm}
\caption{\textit{Left panel}: The HMI continuum image with the yellow contour (level= 22000) showing the boundary between the umbra and penumbra, the black contour (level= 55000) demarcating the boundary between the penumbra and the quiet sun (typically $\sim$ 62000). The blue contours are the fan loops as seen in the AIA 171~{\AA} channel (defined by levels between 800 to 1800 whereas the quiet region has typical values of 100-200). \textit{Middle panel}: Intensity map of AIA 171~{\AA} (blue contours) over-plotted on IRIS-SJI \ion{Si}{4} 1400~{\AA}. The superimposed cyan box denotes a portion of the IRIS raster image showing that the fan loops originate from the footpoint region identified with IRIS. The black contour denotes the sunspot penumbra (level= 55000) whereas the yellow contour denotes the sunspot umbra (level= 22000). \textit{Right panel}: A blow-up of the AIA 171~{\AA} channel image displaying the fan loops and where they are rooted in the sunspot umbra (denoted by the outer red contour). The 
inner red contour represents a region of anomalous polarity within the sunspot umbra.}\label{location}
\end{figure*}

In order to have a clear understanding of the location of footpoints of fan loops with respect to the sunspot, in the left panel of Figure~\ref{location} we show the HMI continuum image over-plotted with blue contours of fan loops obtained from AIA 171~{\AA}. The yellow (level = 22000) and black (levels=55000) contours demarcate the boundary of umbra and penumbra of the sunspot respectively. The middle panel shows IRIS 1400~{\AA} slit-jaw image over-plotted with same contours as in left panel. The overlying cyan box indicates the IRIS raster FOV (reduced along y-direction). The right panel image shows AIA~171{\AA} image over-plotted with B-field contours of level -1200 G, demarcating the big sunspot as well as the anomalous polarity region inside the umbra. As can be inferred from the images, the fan loops are rooted well inside the umbra, exactly at the location where sunspot shows anomalous behaviour. Additionally, the footpoints appear brighter in the IRIS \ion{Si}{4} 1400~{\AA} slit-jaw image. We have 
checked for the HMI magnetogram data and found that the region with anomalous magnetic field persisted over few days. So did the fan loops.


Since we are interested in quiescent AR fan loops, it is important to demonstrate that the structures do not show any significant change during the time of the observations. Therefore, we look at the light curves along the fan loops and check for their stability over time. The fan loops emerging from the sunspot region in the AIA 171~{\AA} channel and the corresponding variation of intensity along a loop over the entire duration of all six rasters of IRIS ($\sim$32 minutes between 10:31:15 UT and 11:03:07 UT) shows that the maximum fluctuation over the entire period is $<4\%$, near the apex. However, the fluctuations closer to the footpoints are smaller. Note that the fluctuations are computed as the difference of maximum and minimum intensities over the mean intensity over all the six rasters. It is emphasized that no treatment of background/foreground intensities was performed. According to \citet{FueK_2015}, any small intensity fluctuation in loops could be attributed to the background/foreground 
contribution. From the temporal evolution plot as well as visual inspection of the IRIS and AIA movies, we are confident that there has been no major eruptive event within this duration that could produce such foreground/background changes.

\subsection{Measurement of Electron Density}

Using the density sensitive line pairs of \ion{O}{4} (IRIS) (1399.77~{\AA} and 1401.16~{\AA} formed at $\log\,[T/K] = 5.15$) and \ion{Si}{10} (EIS) (258.37~{\AA} and 261.04~{\AA} formed at $\log\,[T/K] = 6.15$), we have determined the electron densities at the footpoints and in the fan loops. 

\begin{figure*}
\centering
\includegraphics[width=0.95\textwidth]{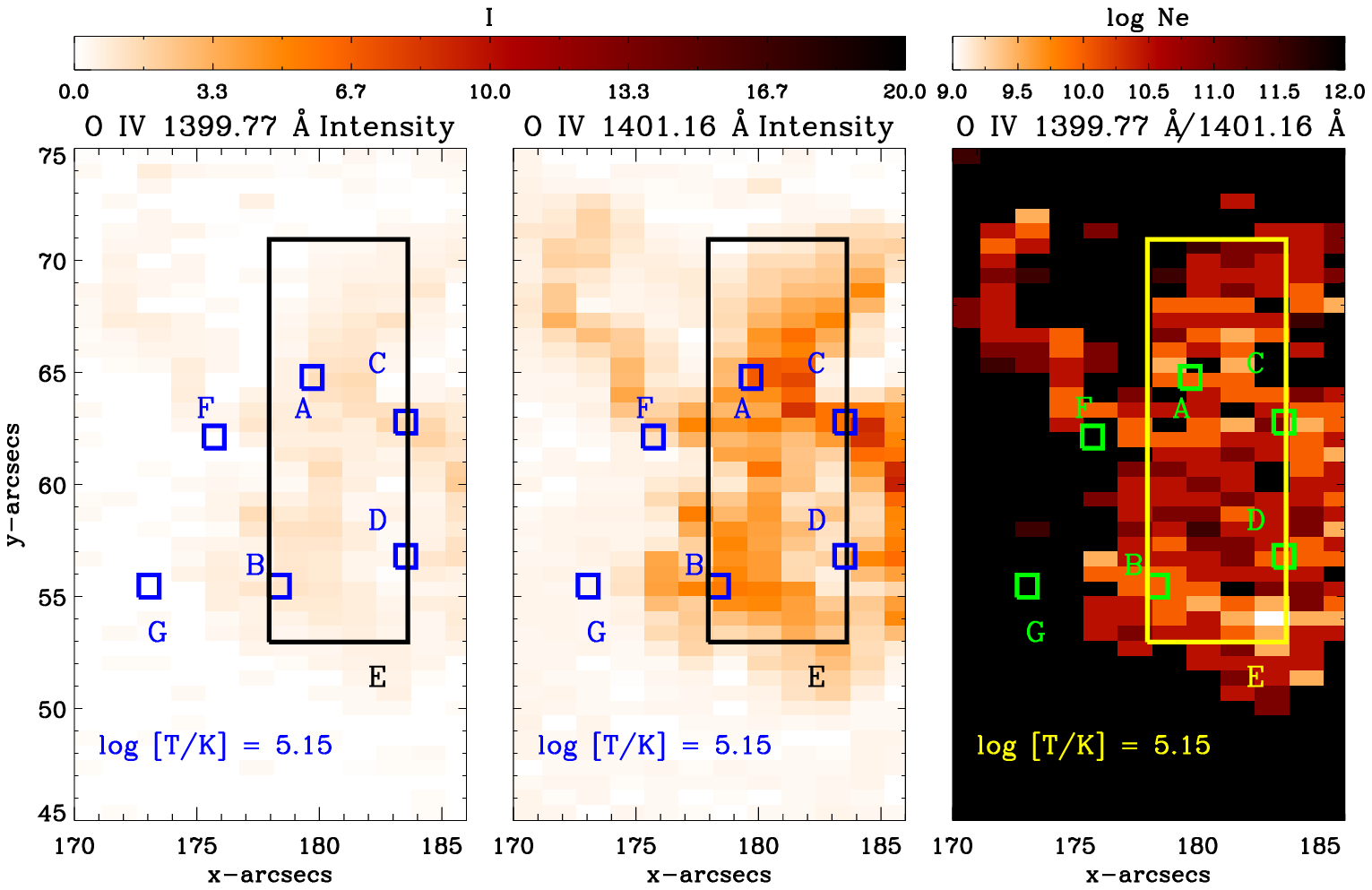}
\vspace{-5mm}
\caption{Intensities in the two \ion{O}{4} lines of IRIS formed at $\log\,[T/K]=5.15$ (left and middle panel) and density (right panel) map obtained from them. Since the \ion{O}{4} lines are very weak, 4x4 pixel binning has been carried out.} \label{density}
\end{figure*}
\begin{table}
\centering
\caption{Average electron densities in four small boxes (Figure~\ref{density}) within the footpoints of fan loops investigated using IRIS \ion{O}{4} and EIS \ion{Si}{10} line pairs. A 20$\%$ uncertainity is acceptable on these figures.}\label{density_table}
 \begin{tabular}{| c | c | c |} \hline
Location   &log $N_{e}\pm 20\%$  &log $N_{e}\pm 20\%$\\ 
            & (\ion{O}{4}) &(\ion{Si}{10})\\ \hline
A &10.0 &9.1\\
B &9.9 &8.9\\
C &10.4 &9.0\\  
D &9.9 &8.8\\    \hline
E &10.1 &8.9\\ \hline
\end{tabular}  
\end{table}

The aim is to compute the average electron density at the footpoint of fan loops, denoted by box E (5.6\arcsec x 17.84\arcsec) in all three panels of Figure~\ref{density}. Figure~\ref{density} provides the two intensity maps obtained for \ion{O}{4} lines (left and middle panel) and the derived density map (right panel). Note that the FOV in the y-direction has been reduced so as to zoom into the footpoints in all the IRIS maps shown in the paper. \ion{O}{4} lines, being weak have fitting problems at pixels with poor counts. In order to improve the SNR, we binned the data by 4 $\times$ 4 pixels. 

In order to scrutinise the goodness of the fits, we randomly pick six small regions (e.g., A-D (at the footpoints) and F \& G (away from footpoints), each being 0.66\arcsec $\times$ 0.83\arcsec shown in Figure~\ref{density}) scattered over the IRIS raster FOV. The fitting has worked very well in the footpoint regions (i.e., in regions A-D) but not in the other regions (i.e. in F and G). In addition, we have worked out the average density in fan loops by considering a bigger box E. The average densities obtained in the four small boxes (A-D) and the big box (E) are given in Table~\ref{density_table}. Including the three factors that incorporate errors in the density estimation - photon count error, fitting error and atomic data errors, we reckon that the total uncertainty in the measurement should not exceed 20$\%$ of the estimated values. On an average, the density at the footpoints of the fan loops within box E is estimated to be $\log\,N_{e} \sim$ 10.1 cm$^{-3}$. For the boxes A to E, shown in 
Figure~\ref{density}, we have also estimated the electron densities using the \ion{Si}{10} line pair observed by EIS. Note that the data for the \ion{Si}{10} lines have been binned by 4 pixels in the y-direction in order to increase the SNR. The densities obtained are also listed in Table~\ref{density_table}. 

The values of electron densities given in Table~\ref{density_table} reveal that the densities measured using \ion{Si}{10} at $\log\,[T/K]=6.15$ is lower than those measured using \ion{O}{4} at $\log\,[T/K]=5.15$. This is suggestive of constant pressure at the footpoints of fan loops. In addition, it also suggests that there are probably a number of coronal strands within the volume where these densities are measured. The plasma in some of these strands is at $\log\,[T/K]=5.15$ and for some others is at $\log\,[T/K]=6.15$. This could be better confirmed by estimating the spectroscopic filling factor \citep{CarK_1997} that requires the structures to be resolved, for example as in \cite{TriMD_2009,GupTM_2015}. Unfortunately, the structures at the footpoints in the present study are not very well resolved, which prohibited us from doing such estimates. We further note that due to poor counts, the estimate of \ion{Si}{10} densities likely suffers from a large uncertainty and may be considered as an upper limit.

\subsection{Temperature structure of fan loops}

\begin{figure}
\centering
\includegraphics[width= 1.2\linewidth,left]{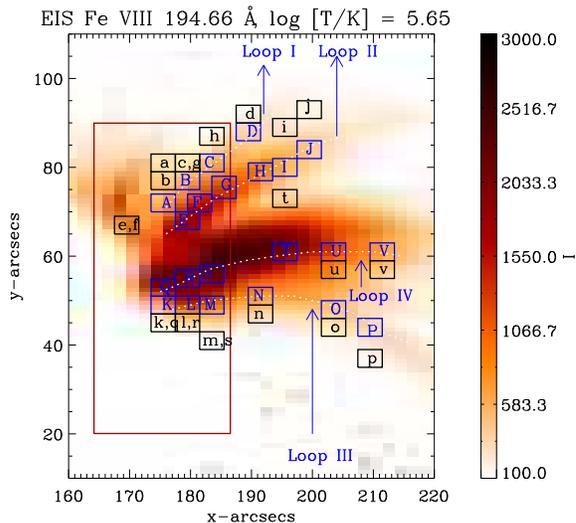}
\caption{EIS \ion{Fe}{8} 194.66~{\AA} image showing the fan loops. Four loops (Loop I, II, III and IV) have been identified. The boxes (3$\arcsec$ x 4$\arcsec$) mark the regions selected for Emission Measure studies. Capital letters are used to identify those sampling the loops whereas those indicated with small letters sample the respective backgrounds. The brown box outlines the IRIS raster FOV, which essentially captures the footpoints of the fan loops. }\label{em_loop}
\end{figure}

Since we have fan loop observations across a range of temperatures, we have produced EM-Loci \citep{JorW_1971, DelLM_2002} plots for four loop structures (loops I, II, III and IV as indicated in Figure~\ref{em_loop}) in order to follow the temperature structure along the loop length. Several small boxes (A-V, each 3\arcsec x 4\arcsec) have been identified on these loops. The numbering with capital letters identify the regions on the loops whereas the numbering with small letters indicates the respective background/foreground. A radiometric calibration has been performed on the IRIS spectral data using the IRIS software\footnote{ITN 26: A User’s Guide to IRIS Data Retrieval, Reduction and Analysis, September 2015}. For the calculation of the contribution function, photospheric abundances\footnote{$sun\_photospheric\_2011\_caffau.abund$} and ionization equilibrium\footnote{$chianti.ioneq$} given by CHIANTI (v7.1.3) spectral synthesis package are used \citep{DerML_1996, LanYD_2013}.
\begin{figure*}
\centering
\includegraphics[width=0.95\textwidth]{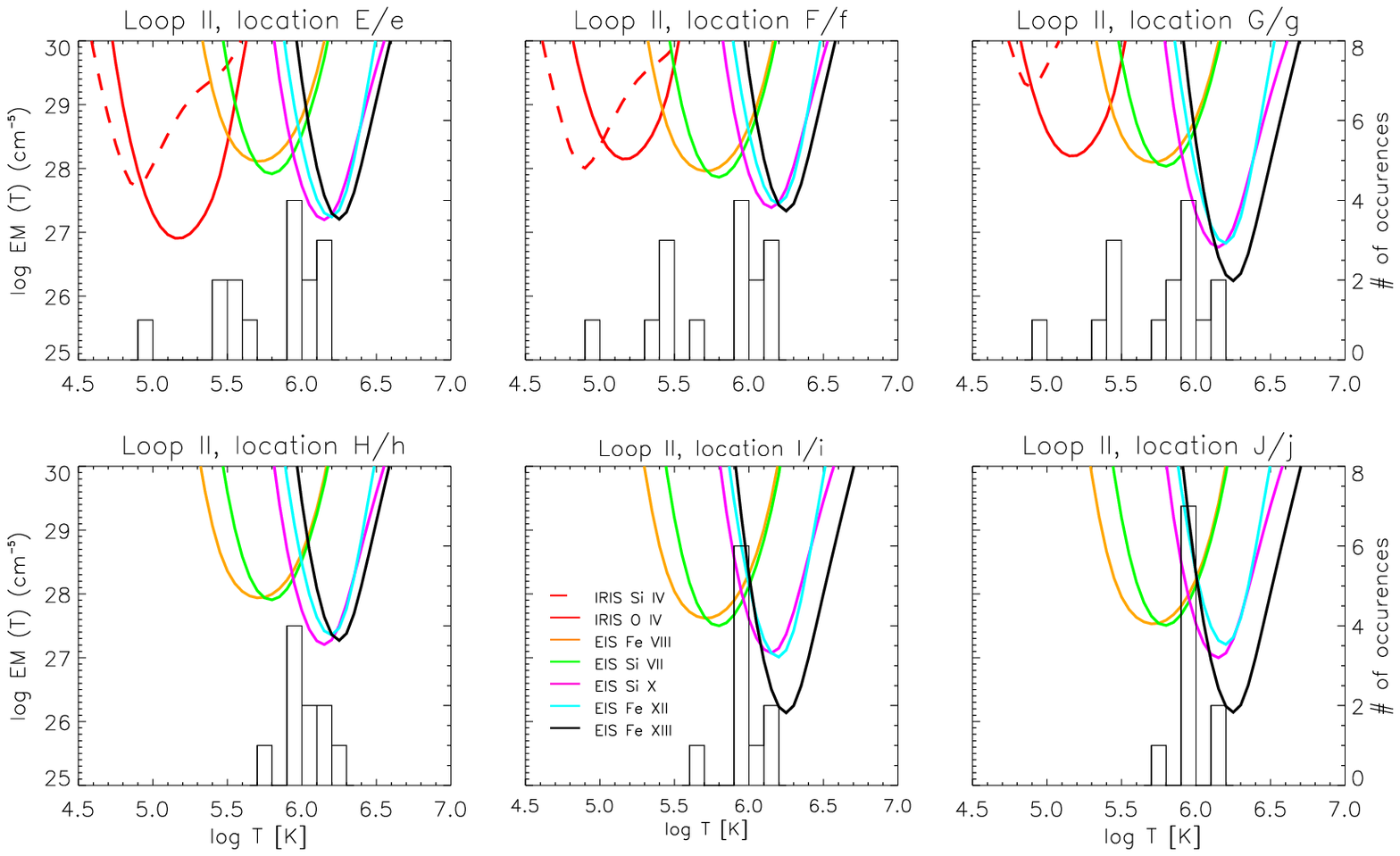}
\vspace{-15mm}
\caption{Emission Measure (EM) loci plots along loop II (Figure~\ref{em_loop}) using IRIS and EIS lines. The histograms plotted at the bottom of each panel represents that number of crossings within a temperature bin of $\log\,[T/MK]=0.1$. The lines to which the plotted curves correspond to are labelled in the second panel of the bottom row. The solid and dashed red lines are for the IRIS lines (\ion{O}{4} and \ion{Si}{4}) respectively. The solid lines of other colors denote the EIS lines as shown in the figure itself.}\label{em_loci2}
\end{figure*}

We have obtained the EM-Loci curves for all the four loops (Figure~\ref{em_loop}) but shown in Figure~\ref{em_loci2} the EM-Loci plot for loop II only. We emphasise here that the other loops provide very similar results. They are not shown here for brevity. The boxes far away from the footpoints have no signatures of the low temperature lines (\ion{Si}{4} and \ion{O}{4}) observed by IRIS. The IRIS lines are available only at the first three locations (first three panels, upper row). Each panel corresponds to a pair of locations (indicated at the top of each panel). The intensity within the region denoted by the capital letter represent the loop whereas the region denoted by the corresponding small letter is considered as the background/foreground. The lines to which the plotted curves correspond to are labelled in the second panel of the bottom row. From Figure~\ref{em_loci2}, we note that closer to the footpoints, the EM-Loci curves for the IRIS lines intersect at one point ($\log\,[T/K]\sim4.95$) and those 
for the EIS spectral lines intersect at another point ($\log\,[T/K]\sim5.95$). This difference between IRIS and EIS could be due to EIS and IRIS cross-calibration. However, the difference is rather too large to be explained just by considering cross calibration. Another possibility could be that the plasma in the fan loops has two temperature components {--} a cooler component (seen by IRIS lines) and another, warmer component (observed by EIS lines) {--} at the footpoints. Considering the fact that we have obtained two different values of electron densities at two temperatures in the previous section, the existence of two plasma components seems more likely.

At the bottom of each panel, the histograms are shown which indicate the number of crossings in each temperature bin of width $\log\,[T/MK]=$ 0.1.  We have defined that the formation temperature of the fan loops is the middle point of the temperature bin where at least four such crossings are present. Going by that convention, it is noted that the maximum number of lines cross within the bin $\log\,[T/K]=$ 5.90 to 6.00 at all the six locations of loop II $i.e.$ the temperature of fan loops is $\log\,[T/K]=$ 5.95, which is similar to the values obtained by \citet{BroWY_2011}. The errors are estimated to be one bin on either side of the bin with maximum number of crossings. The plots also reveal that the temperature remains almost constant at $\log\,[T/K]\sim5.90$ to 6.0 along the lengths of the loops. From the AIA and EIS intensity maps (Figure~\ref{aia_context}), it is seen that the loops are most prominent in the AIA 171~{\AA} channel and in the \ion{Fe}{8} and \ion{Si}{7} spectral lines (all of these have 
a peak formation temperature around $\log\,[T/K]=5.80$). This supports the deduced temperatures.

\subsection{Measurement of Doppler Shift}
\begin{figure}
\centering
\includegraphics[width=1.5\linewidth]{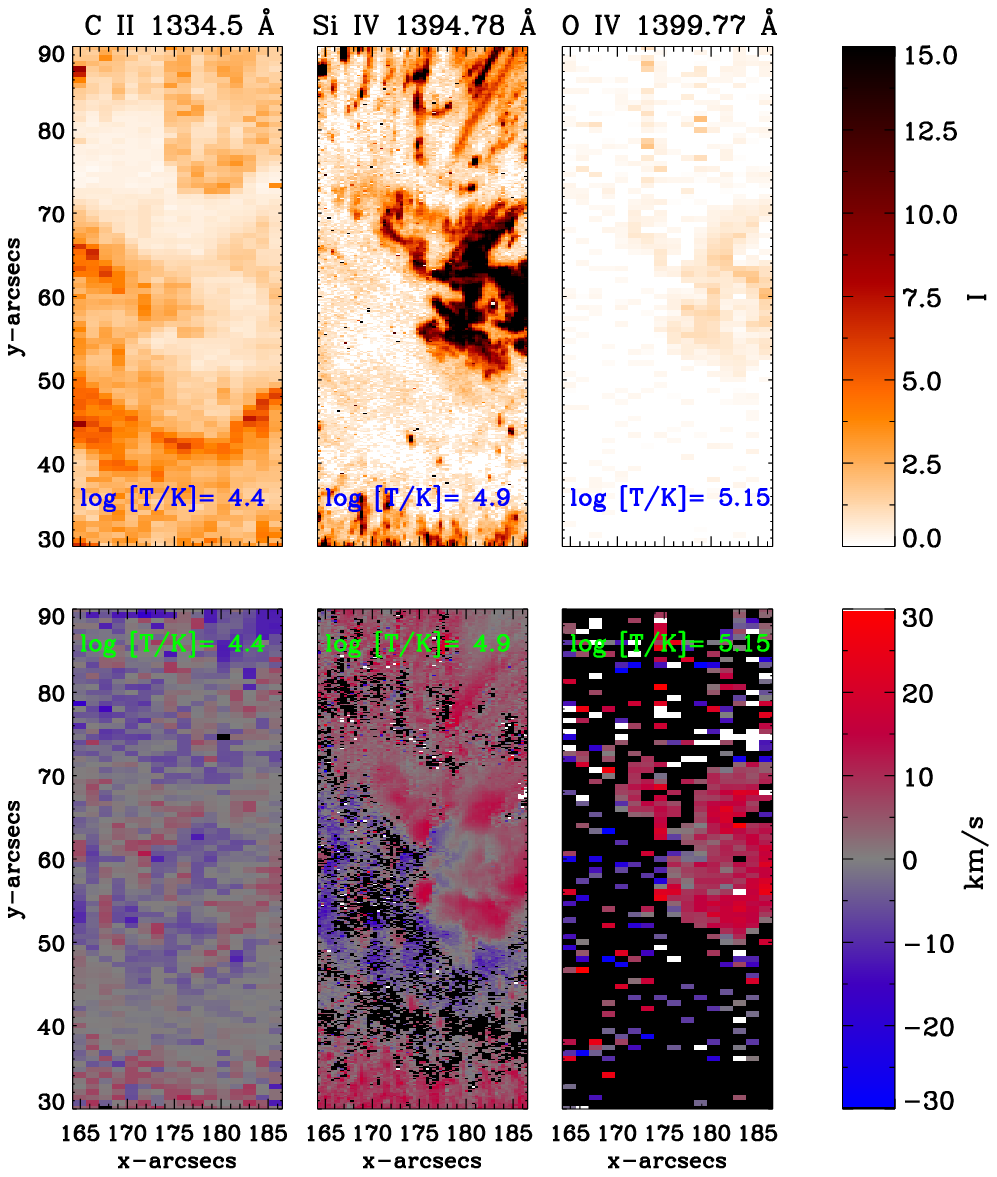}
\caption{Monochromatic intensity (plotted as negatives) and Doppler maps obtained in \ion{C}{2} 1334.5~{\AA} ($\log\,[T/K]=4.40$), \ion{Si}{4} 1394.78~{\AA} ($\log\,[T/K]=4.90$) and \ion{O}{4} 1399.77~{\AA} ($\log\,[T/K]=5.15$) lines observed with IRIS. The images have been arranged in order of increasing temperature. The \ion{C}{2} and \ion{O}{4} lines have been 4x4 pixel binned to improve the signal-to-noise ratio.}\label{iris_map1}
\end{figure}

The IRIS spectral data provides us with an opportunity to study the plasma flows at the footpoints of fan loops at transition region temperatures ($\log\,[T/K] =4.40$ to 5.15). Here, we have obtained the intensity and velocity maps in the \ion{C}{2} ($\log\,[T/K]=4.40$), \ion{Si}{4} ($\log\,[T/K]=4.90$) and \ion{O}{4} ($\log\,[T/K]=5.15$) lines observed by an IRIS raster commencing at 10:31:15~UT. It is known that \ion{C}{2} lines may show double peaked profiles at certain locations \citep{RatCL_2015}. Our analysis of the line profiles of \ion{C}{2} lines at the footpoint of fan loops suggest that they could be well represented by a single Gaussian. We note here that there are two lines for \ion{C}{2}, two for \ion{Si}{4} and two for \ion{O}{4} as mentioned in Table~\ref{spectral_lines}. We have derived the intensity and Doppler maps in all six lines but show the results for one spectral line for each ion. The results for the other lines are similar.

The intensity and corresponding Doppler maps for the IRIS lines are shown in Figure~\ref{iris_map1}. Note that the intensity maps are shown in negative. Since the \ion{C}{2} and \ion{O}{4} lines are weak, these have been binned over $4\times4$ pixels. The footpoints of the fan loops are clearly visible in \ion{Si}{4} as well as in \ion{O}{4} lines and are predominantly redshifted. The redshift is weakest in \ion{C}{2} ($\sim$2{--}3~km~s$^{-1}$) with a peak formation temperature at $\log\,[T/K]=4.40$. With increasing temperature it increases to about 10-15~km~s$^{-1}$ in \ion{Si}{4} ($\log\,[T/K]=4.90$) and further increases to $~$15{--}20~km~s$^{-1}$ in \ion{O}{4} ($\log\,[T/K]=5.15$). Note that the average errors in these measurements are about 3~km~s$^{-1}$.


The footpoint region has been rastered by IRIS six times over a period of $\sim$32 minutes. This provided us with an opportunity to study the variation of Doppler shifts in IRIS lines as a function of time. We have chosen a region, which covered the entire footpoint to study the variation of the average Doppler shift. It is seen that the velocities are relatively stable with a tendency towards a decreasing strength of downflows.
 
\section{Summary and discussion}\label{con}

In this paper, we have studied the plasma parameters of fan loops (at the footpoints as well as along the loops) using observations recorded by IRIS, EIS and AIA. The spectroscopic observations were used to measure the parameters (electron density, temperature and Doppler shifts), whereas the high cadence imaging observations provided by AIA were used to make sure that the loops were not evolving drastically during the course of the IRIS and EIS raster observation. In addition, AIA data were used to coalign the observations from IRIS and EIS. 

The fan loops are observed at near-simultaneous times by AIA, EIS and IRIS. The footpoints of fan loops are seen at both chromospheric and transition region temperatures covering $\log\,[T/K]=$ 4.90 to 5.15. At upper transition region temperatures ($\log\,[T/K]=5.65$ and 5.80), the main body of the loops are distinctly visible. They emanate from the footpoints rooted inside the umbra of a sunspot and end at an unknown location far away (Figure~\ref{aia_context}). They become somewhat less  discernible (more diffuse) as temperature increases, similar to warm loops \citep[][]{TriMD_2009, GuaRP_2010}. We emphasise here that the footpoint, which is rooted inside the umbra of the sunspot, shows anomalous behaviour in magnetic field measurements. 

Below we have summarised the main results obtained in this study:

\begin{enumerate}
	\item Electron densities in various regions at the footpoints of the fan loops are measured using the density sensitive line pairs of \ion{O}{4} ($\log\,[T/K]=5.15$) and \ion{Si}{10} ($\log\,[T/K]=6.15$) observed with IRIS and EIS respectively. The average electron density at the footpoints of fan loops is $\log\, N_{e}=10.1~cm^{-3}$ for \ion{O}{4} and $\log\, N_{e}=8.9~cm^{-3}$ for \ion{Si}{10}.

	\item The temperature structure in the loops (cross-field as well as along the loops) was studied using the EM-Loci of the spectral lines observed both with IRIS and EIS. For this purpose, various locations along four different loops were selected. The locations adjacent to the loops were considered as the background (see Figure~\ref{em_loop}). The IRIS lines are only visible close to the footpoints of the loops. Based on these measurements, we find that there are two components of the plasma (at the footpoints the average temperature from IRIS lines is $\log\,[T/K]\sim 4.95$ and from EIS lines it is $\log\,[T/K] \sim 5.95$) and remains constant thereafter. In the upper part of the loops, the EM loci curves from EIS suggest that all four loops studied are mildly multi-thermal  across the line-of-sight around $\log\,[T/K]=5.95$.
	

	\item The Doppler velocities of the plasma at the footpoints of fan loops are studied using spectral data from IRIS in \ion{C}{2} ($\log\, [T/K] = 4.40$), \ion{Si}{4} ($\log\, [T/K] = 4.90$) and \ion{O}{4} ($\log\, [T/K] = 5.15$). In all these lines the plasma inside the fan loops is predominantly redshifted (downflow) by 2{--3}~km~s$^{-1}$, 10{--}15~km~s$^{-1}$ and 15{--}20~km~s$^{-1}$ respectively and that it increases with increasing temperature, within the observed temperature range. We further note that the observed redshifts at the footpoints persist for a span of more than 30 minutes.

\end{enumerate}

Our measurements of electron densities being higher at lower temperatures and vice-versa suggest that the fan loops are at constant pressure. It further suggests that the loops are comprised of several loop strands within the volume being studied. This is further verified by the temperature structure obtained using EM-loci analysis that shows that there are two component plasmas at the footpoints, one being detected in cooler lines observed by IRIS and other by EIS lines. The Doppler measurement also show that there is plasma at the temperature which is detected by the temperature analysis. Unfortunately, we did not have good enough wavelength calibration to derive velocities at higher temperatures using EIS lines. 

A comprehensive understanding of the physical parameters provides important constraints on the modelling of active region loops. In addition, the observed patterns of density, temperature and flows can be compared with those predicted by different models. In general, there are two mechanisms put forward to explain the heating of active region loops - i.e. high frequency nanoflares (steady heating where thermal conduction flux is eventually balanced by the radiative output) and low frequency nanoflares (impulsive heating) where the enthalpy flux \citep[see e.g.,][]{BraC_2010} from the corona is maintained by the radiative cooling. The high and low frequency scenarios are defined as how frequently the heating occurs as compared to the time taken for the cooling of the loops, once a heating event has taken place. If this interval between two consecutive heating events is small as compared to the cooling time then it is defined as high-frequency heating, thereby allowing a minimum loss of energy between those 
two events (for further explanation see e.g.,\cite{TriKM_2011}). The observational signatures for high frequency heating are: narrow EM distributions ($i.e.$ isothermal cross-field structures) and no Doppler motion unless the loops are asymmetric \citep{BorM_1982,MarB_1983,MarWX_2004}. For low frequency heating, the signatures are instead multi-thermal structures across the loops coupled with Doppler motions. However, the width of the EM curve can vary depending on the nature of nanoflare storms for impulsive heating \citep[][]{Kli_2006,Car_2014}. 

At the footpoint of fan loops our observations show that the plasma is at least a two component thermal structure. At greater heights the temperature across the loops becomes mildly multi-thermal. At the fan loop footpoints, the plasma is predominantly redshifted, that increases with increasing temperature within the observed temperature range ($\log\,[T/K] = 4.40$ to 5.15). The observed temperature structures and Doppler patterns are in agreement with the prediction from low frequency nanoflares and point towards the interpretation that the fan loops are heated via impulsive heating mechanism. However, this is entirely valid only if the loops are symmetric. In the case of asymmetric loops, there tend to be Doppler motions in the plasma due to differences in pressure as was shown by \citet{MarB_1983}. However, the Doppler shifts introduced due such asymmetries are way smaller ($\sim$ 4{--}5~km~$s^{-1}$) than those observed in the current study (15{--}20~km~$s^{-1}$). Therefore, it is plausible to rule out 
that the flows that are being observed here is entirely due to the geometrical asymmetries. 

One of the most important inferences of impulsive heating mechanism is that the plasma goes through sufficient cooling and draining before getting re-heated, implying that all the plasma that leaves the corona must pass through a range of transition region temperatures \citep[see][]{Car_1994}. For a loop having  constant pressure with time and that goes through cooling, \cite{Car_1994} showed that the speed of the plasma flow in the transition region can be approximated as V$_{T}$ $\sim$ $\frac{T_{T}}{T_{c}}$ $\frac{L}{\tau_{r}}$, where T$_T$ and T$_c$ are the transition region and coronal temperatures respectively, L is the loop half length and $\tau_{r}$ is the radiative cooling time. For a projected loop half length of $\sim$100~Mm (as estimated in the current study), and a typical radiative cooling time of 500{--}2000~s, we find that for a coronal temperature of $\log\,[T/K] = 5.95$ (as deduced from EM-Loci analysis), the Doppler shifts in the spectral lines of \ion{C}{2}, \ion{Si}{4} and \ion{O}{4} 
should be in the range 1.5{--}5.5 (observed values are 2-3) km~s$^{-1}$, 4.5{--}18 (observed values are 10-15) km~s$^{-1}$ and 8{--}31 (observed values are 15-20) km~s$^{-1}$ respectively. These values are within the observed limits in the current study. More such observations and further modelling are required to reach a firm conclusion. The results obtained here provides further constraints and inputs for modelling of active region fan loops.

\acknowledgements{We sincerely thank the referee Prof Peter Cargill for his constructive comments that have improved the paper. We also thank Hardi Peter, Giulio Del Zanna, and Hui Tian  for their helpful inputs regarding EIS, IRIS observations and J.~M.~Borrero for discussion on HMI anomalies, respectively. AG is funded by the Max Planck partner group of MPS at IUCAA. DT acknowledges the Max Planck partner group of MPS at IUCAA. This work was also partly supported by the BK21 plus program through the National Research Foundation (NRF) funded by the Ministry of Education of Korea. GRG is supported through the INSPIRE Faculty Award of Department of Science and Technology (DST), India. HEM acknowledges the support of STFC. We thank the IRIS, EIS and SDO consortium for their open data policy. CHIANTI is a collaborative project involving George Mason University, the University of Michigan (USA) and the University of Cambridge (UK).}

\end{document}